\documentclass[twocolumn,prl,aps,superscriptaddress]{revtex4}
\usepackage{amssymb}
\usepackage{cancel}
\usepackage{color,graphicx}
\usepackage{amsmath}
\usepackage{amsbsy}
\usepackage{amsthm}
\usepackage{bbm}
\usepackage{epsfig}
\usepackage{lscape}
\usepackage{float}
\usepackage{graphicx}
\font\zz=cmbxti10

\usepackage[colorlinks=true,citecolor=blue,linkcolor=black,urlcolor=black]{hyperref}
\usepackage{bbm}

\usepackage{subfigure}
\usepackage{dcolumn}
\usepackage{color,epstopdf}
\usepackage{amscd}
\usepackage{amsfonts}  
\usepackage[]{amsmath}
\usepackage{amssymb}    
\usepackage{mathrsfs}
\usepackage{verbatim}
\usepackage[]{cases}
\usepackage{amsmath}
\usepackage{wasysym}
\usepackage[utf8]{inputenc}
\usepackage[T1]{fontenc}
\usepackage{mathtools}
\usepackage{todonotes}

\newcommand{\ket}[1]{\left| {#1} \right\rangle}
\newcommand{\bra}[1]{\left\langle {#1}\right|}

\newcommand{\1}{\mathbbm{1}}

\begin{document}
	
\title{Linear-response approach to critical quantum many-body systems}

\author{Ricardo Puebla}
\affiliation{Instituto de F{\'i}sica Fundamental, IFF-CSIC, Calle Serrano 113b, 28006 Madrid, Spain}
\affiliation{Centre for Theoretical Atomic, Molecular, and Optical Physics, School of Mathematics and Physics, Queen’s University, Belfast BT7 1NN, United Kingdom}
 \author{Alessio Belenchia}
\affiliation{Institut f\"{u}r Theoretische Physik, Eberhard-Karls-Universit\"{a}t T\"{u}bingen, 72076 T\"{u}bingen, Germany}
\affiliation{Centre for Theoretical Atomic, Molecular, and Optical Physics, School of Mathematics and Physics, Queen’s University, Belfast BT7 1NN, United Kingdom}
 \author{Giulio Gasbarri}
 \affiliation{Department of Physics and Astronomy, University of Southampton, Highfield Campus, SO17 1BJ, United Kingdom}
 \affiliation{F\'isica Te\`orica: Informaci\'o i Fen\`omens Qu\`antics, Department de F\'isica, Universitat Aut\`onoma de Barcelona, 08193 Bellaterra (Barcelona), Spain}
  \author{Eric Lutz}
\affiliation{Institute for Theoretical Physics I, University of Stuttgart, D-70550 Stuttgart, Germany}
 \author{Mauro Paternostro}
\affiliation{Centre for Theoretical Atomic, Molecular, and Optical Physics, School of Mathematics and Physics, Queen’s University, Belfast BT7 1NN, United Kingdom}

\begin{abstract}
The characterization of quantum critical phenomena is pivotal for the understanding and harnessing of quantum many-body physics. However, their complexity makes the inference of such fundamental processes difficult. Thus, efficient and experimentally non-demanding methods for their diagnosis are strongly desired. Here, we introduce a general scheme, based on the combination of finite-size scaling and the linear response of a given observable to a time-dependent perturbation, to efficiently extract the energy gaps to the lowest excited states of the system, and thus infer its dynamical critical exponents. Remarkably, the scheme is able to tackle both 
integrable and non-integrable models, prepared away from their ground states. It thus holds the potential to embody a valuable diagnostic tool for experimentally significant problems in quantum many-body physics. 
 \end{abstract}
	
\maketitle

The investigation of quantum many-body systems plays a pivotal role in our understanding of  novel phases of matter, both in- and out-of-equilibrium~\cite{son97,voj03,Sachdev,Dutta,Polkovnikov:11,Eisert:15}. Important applications include quantum information theory~\cite{ami08} and material science~\cite{ful14}.
One of most puzzling aspects  of such systems are quantum phase transitions (QPT)~\cite{son97,voj03,Sachdev,Dutta}. In contrast to their classical counterparts, which stem from classical thermal fluctuations, they occur at zero temperature in energy eigenstates of interacting quantum many-body systems as an external non-thermal parameter is varied, and are thus driven by quantum fluctuations. Similarly to their classical counterparts, continuous quantum phase transitions can  be classified according to universality classes featuring the same critical exponents~\cite{son97,voj03,Sachdev,Dutta}. As a consequence, distinct quantum many-particle systems   belonging to the same universality class will display equivalent critical properties, independently of their microscopic details. 

The determination of the critical exponents of a QPT is a major theoretical and experimental challenge~\cite{son97,voj03,Sachdev,Dutta} that, for continuous classical phase transitions, has been addressed by examining the behavior of  thermodynamic response coefficients, such as susceptibilities, compressibilities and  heat capacities~\cite{ma76}. Other approaches have been developed over the years, including the study of the response of information-theoretic quantities such as quantum correlations~\cite{ami08,DeChiara18} and state fidelity~\cite{gu10}, and the tracking of the behavior of geometric phases~\cite{car20}. All such approaches pose significant difficulties that make the availability of experiment-ready techniques for the inference of the critical exponents of a given transition a pressing need. 

A potentially fruitful avenue is provided by linear response theory, a versatile tool of statistical mechanics for the investigation of (non-)equilibrium complex systems, from hydrodynamics to condensed-matter physics, that connects  the equilibrium fluctuations of a classical or quantum system to its response to weak perturbations~\cite{kubo1957statistical,kubo1966fluctuation,hanggi1982stochastic,marconi2008fluctuation,Naze22}. The linear response formalism, which has recently been further extended to non-equilibrium steady-states~\cite{Prost09,bai09,sei10,meh18,kon18}, can be used to either predict the behavior of the perturbed system from its known equilibrium properties or, vice versa, to infer its equilibrium properties  from the response to a known perturbation.

Building on such fundamental links between equilibrium features and non-equilibruum response, here we show that dynamical critical exponents of many-body quantum systems undergoing a QPT can be efficiently extracted from the linear response of a suitable observable. We focus on spin systems (and related fermionic models) owing to their central theoretical~\cite{son97,voj03,Sachdev,Dutta} and experimental~\cite{ron05,col10,muk12,Friedenauer:08,Kim:10,Islam:13,Richerme:14,Jurcevic:17,kin14,keesling19,nie20,cai21,eba21} relevance.  We consider two paradigmatic integrable systems, the one-dimensional transverse field Ising model  (TFIM)~\cite{son97,voj03,Sachdev,Dutta} and a long-range Kitaev (LRK) chain of spinless fermions~\cite{Kitaev:01,Vodola:14,Alecce:17}. We find a surprisingly simple relation between linear response following a perturbation and the energy spectrum of the unperturbed system. By combining the linear response after a parameter quench with a finite-size scaling analysis of the energy gap at criticality~\cite{Fisher:72,Fisher:74,Brankov}, we are able to accurately deduce the corresponding dynamical critical exponents. The usefulness of this approach is further highlighted by tackling non-zero temperature initial states and non-integrable models, thus proving its applicability to a range of situations of strong experimental prominence.

\noindent
{\zz Linear response formalism.} We consider a closed quantum system with  Hamiltonian $H_0$ whose ground state is unitarily  perturbed by $\lambda(t) H_1$, where $[H_0,H_1]\neq0$ and $\lambda(t)$ is a small time-dependent parameter. The linear response of a generic observable $B$ of the system, initially prepared in state $\rho_0$, is given by the  Kubo formula~\cite{kubo1957statistical,kubo1966fluctuation,hanggi1982stochastic,marconi2008fluctuation} (we choose units such that $\hbar=1$ throughout the manuscript)
\begin{equation}\label{firstform}
\langle {B}\rangle=\langle {B}\rangle_0+{i}\,\int_{0}^t  \lambda(s)\langle [{B}(t-s),H_1]\rangle_{0}\,ds,
\end{equation}
where ${B}(\tau)=e^{iH_0\tau}B e^{-iH_0\tau}$ 
and $\langle {\cdot} \rangle_{0}$ 
denotes the average over $\rho_{0}$, which may in general describe an equilibrium state or a non-equilibrium steady state \cite{meh18,kon18}. Equation~\eqref{firstform} embodies the starting point of our linear response analysis to quantum phase transitions.

\begin{figure*}[t!]
\centering
\includegraphics[width=\linewidth]{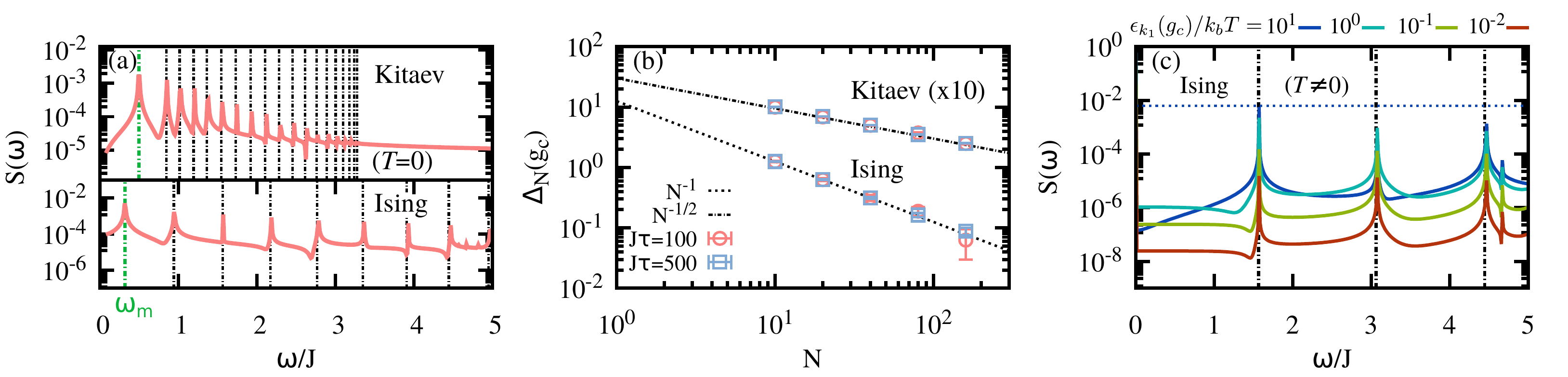}
\caption{\small{(a) {\bf Bottom}: Spectrum $S(\omega)$ of  Eq.~\eqref{eq:MLRsq} for a transverse-field Ising at $g_0=g_c$. {\bf Top}: analogous quantity for Eq.~\eqref{eq:NfLR} in a LRK model for $\mu_0=\mu_c=2J$. In both panels, $\delta\mu=0.01J$ with a total evolution time $J\tau=500$ and $N=40$ particles,  
    with $n_\tau=1000$ evenly sample points in $t\in[0,\tau]$. The dashed lines indicate the exact energy spectrum, with the lowest non-zero frequency $\omega_m$ depicted in green. (b) Finite-size scaling of the energy gap $\Delta_N$ at the critical point for the Ising and LRK model (the latter with $\alpha=5/2$ and $\beta=3/2$). 
For $J\tau=100$ and $n_\tau=100$ (open red circles) and $J\tau=500$ with $n_\tau=1000$ (open blue squares), we find an excellent agreement with the expected respective scalings  $N^{-1}$ and $N^{-1/2}$. A fit to the determined $\Delta_N$ for the Ising yields $z=0.92(5)$ and $z=0.98(2)$, depending on the duration time, close to the exact value $z=1$. For the Kitaev model, the fit results in $z=0.49(1)$ and $z=0.50(1)$, compatible with the expected $z=\beta-1=1/2$ for the chosen $\alpha$ and $\beta$. Similar results can be obtained for other $\alpha$ and $\beta$. (c) Similar to bottom panel in (a) but for initial states with distinct temperature $T$. The horizontal dotted line corresponds to the predicted $S(\omega_m)$ for $\epsilon_{k_1}(g_c)/k_b T=10^1$.}}
\label{fig1}
\end{figure*}

\noindent
{\zz Transverse field Ising model.} In order to illustrate the features of the method that we propose, we address the simple yet informative and relevant example embodied by the transverse-field Ising model with nearest-neighbor interactions. The corresponding Hamiltonian reads~\cite{son97,voj03,Sachdev,Dutta}
\begin{equation}\label{HTFIM}
H_{\rm TFIM}=-J\sum_{j=1}^N \left(g \sigma_j^x +\sigma_j^z\sigma_{j+1}^z\right),
\end{equation}
with $N$ the number of particles of the model, $J>0$ the exchange constant, $g>0$ the coupling parameter, and $\sigma_j^{x,y,z}$ denoting the usual Pauli spin operators. 
In order to fix the ideas and without affecting the generality of our conclusions, we choose $N$ even and periodic boundary conditions for convenience, i.e. $\sigma^{x,y,z}_{N+1}=\sigma^{x,y,z}_{1}$. Equation~\eqref{HTFIM} features a quantum phase transition from an ordered to a disordered paramagnetic phase at $g=g_c=1$ \cite{son97,voj03,Sachdev,Dutta}.

Through Jordan-Wigner and  Fourier transformations, Eq.~\eqref{HTFIM} can be cast in the form of a set of $N/2$ independent Landau-Zener problems $H_{\rm TFIM}=\bigoplus^{N/2}_{n=1} H_{0,k_n}$ for the positive parity subspace with $H_{0,k}=h_k^z\sigma_z^{k}+h_{k}^x\sigma_x^{k}$. Here, we have introduced the quasiparticles operators $\sigma_z^k=\ket{1}\bra{1}_k-\ket{0}\bra{0}_k$ and parameters $h_k^z(g)=2J[g-\cos (k b)]$, $h_k^x=2J\sin (k b)$, which are written in terms of the allowed wavenumbers $k_n=(2n-1)\pi/(N b)$ with $n\in \{1,...,N/2 \}$ and $b$ the spacing between the spins. 
The  diagonalization of each Landau-Zener Hamiltonian $H_{0,k_n}$ leads to the eigenvalues $\epsilon_{k_n}(g)=2|J|\sqrt{g^2+1-2g\cos(k_nb)}$ with corresponding eigenstates \cite{son97,voj03,Sachdev,Dutta}
\begin{equation}
\label{eigen}
\begin{aligned}
\begin{pmatrix}
  \ket{\phi_{k_n,+}(g)}\\
    \ket{\phi_{k_n,-}(g)}
\end{pmatrix}
=   
\begin{pmatrix}
\sin \theta_{k_n}^{(g)}&-\cos\theta_{k_n}^{(g)}\\
\cos \theta_{k_n}^{(g)}&\sin\theta_{k_n}^{(g)}
\end{pmatrix}
\begin{pmatrix}
 \ket{0}_{k_n}\\
  \ket{1}_{k_n}
\end{pmatrix},
\end{aligned}
\end{equation}
where $\theta^{(g)}_{k_n}=-\arctan[(1+\sqrt{1+\zeta^2_{k_n}})/\zeta_{k_n}]$ is a mixing angle in momentum space and $\zeta_{k_n}=h^x_{k_n}(g)/h^z_{k_n}(g)$. 

We now consider the case where the Ising chain, initially in its ground state at $g=g_0$ so that $H_{0}=H_{\rm TFIM}(g_0)$, undergoes a quench $g_0\rightarrow g_0+\delta g(t)$, leading to the perturbation $H_{1}=-J\sum_{j=1}^N  \ \sigma_j^x$ with time-dependent perturbation parameter $\lambda(t)=\delta g(t)$. In terms of the eigenstates in Eq.~\eqref{eigen}, the initial state reads $\rho_0=\ket{\psi_0}\bra{\psi_0}$ with $\ket{\psi_0}=\bigotimes_{k_n} \ket{\phi_{k_n,-}(g_0)}$, while the free dynamics 
is determined by the set of momentum-space Hamiltonians
$ H_{0,k_n}=\epsilon_{k_n}(g_0)\tilde{\sigma}_z^{k_n}$
with $\tilde{\sigma}_z^{k_n}=\ket{\phi_{{k_n},+}}\bra{\phi_{k_n,+}}-\ket{\phi_{k_n,-}}\bra{\phi_{k_n,-}}$, while 
the perturbation  becomes  $H_{1}=\sum_{k_n>0} \Psi_{k_n}^\dagger H_{1,k_n}\Psi_{k_n}$, where we have introduced the momentum-space fermionic modes $\Psi^\dagger_k=(c_k^\dagger,c_{-k})$. The  Hamiltonians $H_{1,k_n}$ are explicitly given by
\begin{equation}
\begin{aligned}
H_{1,k_n}&=2J\sigma_z^{k_n}
=2J \left[\cos(2\theta^{(g_0)}_{k_n})\tilde{\sigma}_z^{k_n}-\sin(2\theta^{(g_0)}_{k_n})\tilde{\sigma}_x^{k_n}\right]
\end{aligned}
\end{equation}
with $\tilde{\sigma}^{k_n}_x=\ket{\phi_{k_n,+}}\bra{\phi_{k_n,-}}+\ket{\phi_{k_n,-}}\bra{\phi_{k_n,+}}$.

From the above  expressions, we see that the linear response of the  many-body system in Eq.~\eqref{firstform} can be expressed as the sum of the linear responses of a single spin system in each of the momentum subspaces in which the Ising model is decoupled. 
Thus, for a generic observable $B=\sum_k \Psi_k^\dagger B_k \Psi_k$ (transformed into the eigenbasis of each subspace $k$), with $B_k=\sum_{j} b_j^k \tilde{\sigma}_j^k$, and  a steady state of the unperturbed dynamics  of the  form $\rho_0^k=(\1+f_z^k\tilde{\sigma}^k_z)/2$, the linear response for each $B_k$ reads~\cite{SM}
\begin{align}\label{eq:LRIsingGenU}
\left< B_k\right>&=f_z^k b_z^k+f_z^k\int_0^t ds \sum_{n,j}2 c_n(s)\big[b_j^k \varepsilon_{nzj}\cos(2(t-s))\nonumber\\&+ b_j^k\delta_{nj}(1-\delta_{zj})\sin(2(t-s))\big],
\end{align}
where $\varepsilon_{n z j}$ is the Levi-Civita symbol. For a perturbation $g_0\rightarrow g_0+\delta g(t)$, the coefficients are explicitly $c_x(t)=-2J\delta g(t)\sin(2\theta^{(g_0)}_k)$, $c_y(t)=0$ and $c_z(t)=2J\delta g(t)\cos(2\theta^{(g_0)}_k)$. At zero temperature,  $f_z^k=-1$ for all $k$. For simplicity, Eq.~\eqref{eq:LRIsingGenU} has been written  assuming unit frequency for each subspace,  so that $H_{0,k}=\tilde\sigma_z^k$. In general, each subspace evolves  at a different evolution rate, given by the interplay of the coefficients $c_n$ of the perturbation and the eigenfrequencies of the free dynamics $\epsilon_k(g_0)$. In that case, one should rescale the coefficients as $c_n\rightarrow c_n/\epsilon_k(g_0)$, and time as $t\rightarrow t_k=\epsilon_k(g_0) t$ in Eq.~\eqref{eq:LRIsingGenU}. 

We proceed by choosing an observable $B$ that is easily accessible experimentally, namely  the magnetization along the $x$-axis, $M_x=(1/N)\sum_{j=1}^N\sigma_j^x$. Due to the translational symmetry of $H_{\rm TFIM}$, $M_x$ corresponds to the single magnetization of any spin in the chain. The observable $M_x$ in the $k$-momentum subspaces simply reads  $M_{x,k_n}=(2/N) [-\cos(2\theta^{(g_0)}_{k_n})\tilde{\sigma}_z^{k_n}+\sin(2\theta^{(g_0)}_{k_n})\tilde{\sigma}_x^{k_n}]$. Adding all the $k_n$-contributions with their corresponding rescaled coefficients and time, we obtain the linear response of the transverse magnetization to the  quench $ \langle M_x\rangle =(2/N)\sum_{k_n>0}{\cal M}_{k_n}$ with 
\begin{eqnarray}\label{eq:MLRsq}
  {\cal M}_{k}=
  \cos( 2\theta^{(g_0)}_{k})+\frac{4J(\delta g)  \sin^2(2\theta^{(g_0)}_k)}{\epsilon_k(g_0)}
  \sin^2 (\epsilon_k(g_0)t) 
\end{eqnarray}
This result establishes a direct link between the linear response of an observable (in this case the transverse magnetization) to a known perturbation, and the properties of the unperturbed many-body system, specifically the energy spectrum $\epsilon_k(g_0)$. The Fourier spectrum of the response will have frequency components at positions $2\epsilon_k(g_0)$ with amplitudes $\propto 1/\epsilon_k(g_0)$. We shall now illustrate how such relationship can be used to determine the dynamical critical exponent $z$ and thus help identify the universality class of  the quantum phase transition.

Let us recall that the energy gap $\Delta_N$ between the ground and excited state for a  system consisting of $N$ elements vanishes at the critical point in the thermodynamic limit $N\rightarrow \infty$. This is a prominent hallmark of a quantum phase transition~\cite{son97,voj03,Sachdev,Dutta}. Finite-size scaling theory at finite $N$ predicts that, at criticality, such gap vanishes as  $\Delta_N(g_c)\propto N^{-z}$, where $z$ is the dynamical critical exponent~\cite{Fisher:72,Fisher:74,Brankov}. The transverse-field Ising model belongs to the universality class with $z=1$. 
In order to extract the value of $z$ from the linear response expression in Eq.~\eqref{eq:MLRsq}, we  first remark that the 
energy gap between ground and first-excited state of the model reads $\Delta_N(g_0)
= 2\epsilon_{k_1}(g_0)$. Therefore, the critical properties of the model may then be obtained by: (i) first sampling $\langle M_x\rangle$ at various times upon a quench at criticality $g_0=g_c$; (ii) then computing the Fourier spectrum of the transverse magnetization $S(\omega)=\int dt \langle M_x\rangle \exp(-i\omega t)$; (iii) finally determining $2\epsilon_{k_1}(g_c)=  \omega_m$ from the value of the lowest non-zero frequency $\omega_m$. By repeating this scheme for different system sizes $N$, one can retrieve the value of the exponent $z$ from the scaling $N^{-z}$ of the energy gap $\Delta_N(g_c)$~\footnote{We also note that the location of the critical point $g_c$ and the critical exponent $\nu$ can be extracted scanning $\Delta_N(g)$ for various $g$ and $N$. Indeed, according to finite-size scaling~\cite{Fisher:72,Fisher:74}, from $g^*(N)={\min_g\Delta_N(g)}$ follows $g^*(N)-g_c\propto N^{-1/\nu}$ that allows to determine $\nu$ and $g_c$.}.
In order to showcase the success of this procedure, in Fig.~\ref{fig1}(a) we have reported $S(\omega)$ for $\delta g/J=0.01$,  $g_0=g_c$, $N=40$ spins and a total evolution time $J\tau=500$. The position of  the lowest non-zero frequency $\omega_m$ is indicated in green. Fig.~\ref{fig1}(b) further displays the finite-size scaling of the energy gap $\Delta_N(g_c)= \omega_m$ at the critical point evaluated  for $J\tau=100$ (open red circles) and $J\tau=500$ (open blue squares). A fit with $N^{-z}$ yields the respective values $z=0.92(5)$ and  $z=0.98(2)$, which are very close to the exact value $z=1$. 

 \noindent
{\zz Long-range Kitaev chain.}
In order to validate the proposed method in a situation offering a richer phenomenology, we consider a LRK chain of $N$ spinless fermions on a lattice with open boundary conditions. The associated Hamiltonian reads~\cite{Kitaev:01,Vodola:14,Alecce:17}
\begin{equation}\label{lrk}
\begin{aligned}
    H_{\rm LRK}&{=-J}\sum_{j=1}^N 
    \left[\sum_{r>0} \big(J_r c_j^\dagger c_{j+r}{+} d_r  c_j c_{j+r}{+}h.c.\big){-\mu}n_j\right]{+}{\cal K} 
    ,
    \end{aligned}
\end{equation}
where $c_j$ and $c_j^\dagger$ are the fermionic annihilation and creation operators with $n_j=c_j^\dagger c_j$, $\mu$ is the chemical potential that controls the quantum phase transition between ferromagnetic and paramagnetic phases, $J>0$ is a scale coefficient,  $J_r$ and $d_r$ are the hopping and pairing strengths, respectively, and we have introduced the constant ${\cal K}=-J\mu N/2$. The parameters of the model are renormalized as $ J_r^\alpha={1}/{N_\alpha r^{\alpha}}$ and $d_r^\beta= {1}/{N_\beta r^{\beta}}$ with $N_\gamma=2\sum_{r=1}^{N/2}r^{-\gamma}$ following Kac's prescription~\cite{Kac:63} to ensure an extensive energy in the thermodynamic limit. The parameters $\alpha,\beta>1$ are the long-range exponents of the interaction, for hopping and pairing, respectively. For short-range interactions ($\alpha,\beta\rightarrow \infty$), there is a quantum phase transition at $\mu_c=2J$ and the model can be mapped exactly onto the transverse field Ising model~\cite{Kitaev:01}. However, for long-range interactions, and depending on the finite values of $\alpha$ and $\beta$, the critical exponents  are modified, and so the universality class to which the model belongs~\cite{Alecce:17,Vodola:14}.
\begin{figure*}[t!]
\centering
\includegraphics[width=\linewidth]{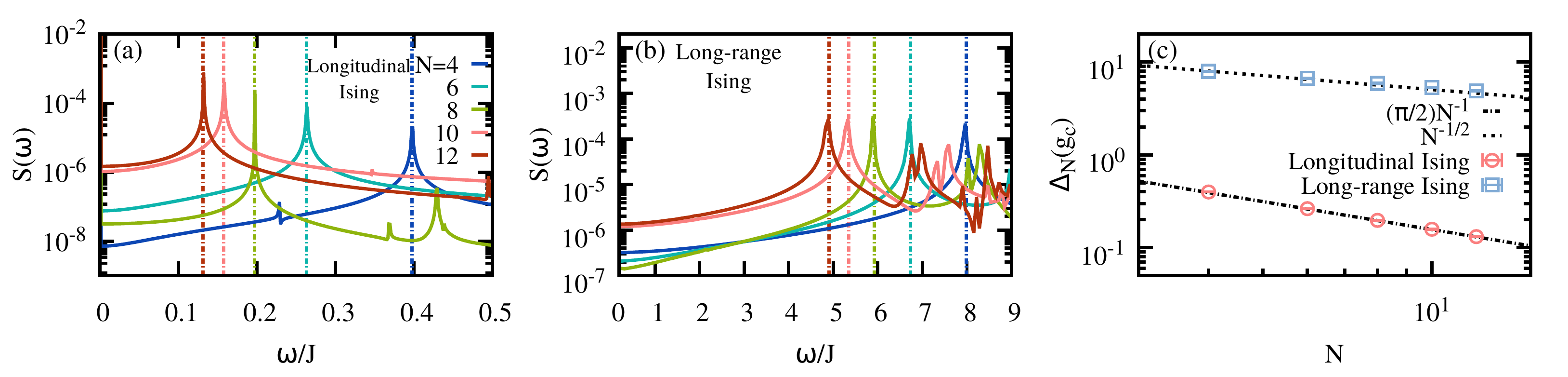}
\caption{\small{(a) Spectrum $S(\omega)$ of the average magnetization  $\langle M_x\rangle$ upon a longitudinal magnetic-field perturbation with $g_0=1$, $\delta h=10^{-3}$ and various sizes $N$. Vertical dashed lines show the position of $\Delta_{\rm s,N}(g_0)$. (b) Spectrum $S(\omega)$ for a transverse-field Ising model with long-range and antiferromagnetic interactions $H_{\rm long}$ with $r=2$ upon a perturbation $\delta g=10^{-2}$ at the critical point $g_c$. (c) Finite-size scaling of the energy gap for both models, which provide $z=1.01(1)$ and $z=0.47(2)$ for the longitudinal and long-range Ising model, respectively.}} 
\label{fig2}
\end{figure*}

The unperturbed Hamiltonian in Eq.~\eqref{lrk} may be diagonalized by taking the Fourier transform of $c_j$ and $c_j^\dagger$~\cite{Vodola:14,Alecce:17}. Using the same notation as for the transverse-field Ising model, one has the momentum-space Hamiltonians   $H_{0,k}=h_z(\mu)\sigma_z^k+h_x\sigma_x^k$ with $h_z(\mu)=\mu/2-2J\sum_{r>0}J_r\cos(kra)/N_\alpha$ and $h_x=-J \sum_{r>0}d_r\sin (kra)/N_\beta$. Upon diagonalization one then  obtains $H_{0,k}=\epsilon_k(\mu)\tilde{\sigma}_z^k$ with $\epsilon_k(\mu)=\sqrt{h_z^2(\mu)+h_x^2}$, and  the mixing angle $\theta^{(\mu)}_k$ akin to Eq.~\eqref{eigen}. The long-range character of the model is encoded in the functions  $h_x$ and $h_z$, which depend on the  parameters $\alpha$ and $\beta$. For $\alpha>1$ and $\alpha<\beta<2$, the dynamical critical exponent is given by $z=\beta-1$~\cite{Defenu:19}, and thus $\Delta_N(\mu_c)\propto N^{1-\beta}$.

In order to study its critical properties, we perturb the LRK chain, prepared in the ground state at $\mu = \mu_0$, by a sudden quench of the chemical potential $\mu_0\rightarrow \mu_0+\delta \mu$, and choose the number of fermions, $N_f=\sum_{j=1}^N n_j=\sum_{k>0} \Psi_k^\dagger \sigma_z^k \Psi_k$, as the observable of interest. We eventually get the  linear response $ \langle N_f \rangle=-\sum_{k>0}{\cal N}_k$ with
\begin{eqnarray}\label{eq:NfLR}
   {\cal N}_k=\cos (2\theta^{(\mu_0)}_k)+\frac{2(\delta \mu) \sin^2(2\theta^{(\mu_0)}_k)}{\epsilon_k(\mu_0)}
   \sin^2(\epsilon_k(\mu_0)t).
\end{eqnarray}
We may determine the critical exponent $z$ by applying the scheme described previously. The top panel of Fig.~\ref{fig1}(a) shows the Fourier spectrum $S(\omega)$ and the position of the lowest non-zero frequency $\omega_m$ (in green) for $\alpha=5/2$, $\beta=3/2$, $\mu_0=\mu_c=2J$, $\delta\mu/J=0.01$ and $J\tau =500$. The fit of  the energy gap $\Delta_N(\mu_c)\propto N^{1-\beta}$ [cf. Fig.~\ref{fig1}(b)] yields the critical exponent  $z=0.49(1)$ (for $J\tau=100$, open red circles) and $z=0.50(1)$ (for $J\tau=1000$, open blue squares). Both values agree very well with the expected  $z=\beta-1=1/2$, further confirming the effectiveness of the linear response approach. 
%

\noindent
{\zz Non-zero temperature.} QPT usually influence a wide portion of the phase diagram of a model, even far from absolute zero~\cite{son97,voj03,Sachdev,Dutta}.  Additionally, the ground state of a many-body system is often difficult to prepare experimentally. It is thus important to be able to detect quantum phase transitions in systems  in  a non-zero temperature  initial state. In this case, the initial unperturbed state of $H_{0,k}$ is a thermal state, $\rho_0^k=(\1+f_z^k\tilde{\sigma}_z^k)/2$ with $f_z^k=-\tanh(\epsilon_k(g_0)/k_bT)$, $T$ the temperature of the system, and $k_b$ the Boltzmann constant. Its linear response thus follows from Eq.~\eqref{eq:LRIsingGenU} and  the energy gap $\Delta_N$ can still be obtained  in a similar manner as for the $T=0$ scenario~\cite{SM}. Fig.~1(c), shows the energy spectrum for the transverse-field Ising model Eq.~\eqref{HTFIM} for various initial temperatures, from which we may again determine the  dynamical critical exponent $z= 1$ with good accuracy.  We note, however, that for temperatures corresponding to energies much larger than the energy gap, the Fourier component at frequency equal to the energy gap $2\epsilon_{k_1}$ is suppressed as $\sim\epsilon_{k_1}/ (k_bT)$ since $f_z^k\rightarrow 0$ as $\epsilon_k/(k_bT)\rightarrow 0$. This sets a boundary of the quantum critical nature of the system at finite temperature as $T\sim |g-g_c|^{z\nu}$~\cite{Sachdev,kin14}. At the critical point, $\Delta_N(g_c)\propto N^{-z}$, and thus the required temperature to resolve the energy gap and the dynamical critical exponent scales as $T\propto N^{-z}$ as the system size increases.

\noindent
{\zz Non-integrable models.} Building on the previous analytical results, we turn our attention to the linear response of non-integrable models. For that we consider a transverse-field Ising model with a longitudinal magnetic field $H_\text{longitudinal}=H_{\rm TFIM}+Jh\sum_{j=1}^N\sigma_j^z$. Choosing again as initial state the ground state of $H_{\rm TFIM}$ at $g_0$ (i.e. $h=0$), a perturbation $\delta h$ breaks the integrability of the model~\cite{Dutta}. As such, the parity $Z_{2}$ ceases to be a conserved quantity, and the state explores the two previously disjoint parity subspaces. For $h=0$, the energy gap between the ground states of these two subspaces is given by $\Delta_{s,N}(g)=E_0^--E_0$ with $E_0^-=-2J(1+\sum_{n=1}^{N/2-1}\sqrt{g^2+1-2g\cos(2n\pi/N)})$. For $g<g_c$, $\Delta_{s,N}(g)$ vanishes exponentially with $N$, while for $g>g_c$ the gap is non-zero and vanishes as $\Delta_{s,N}(g)\propto |g-g_c|^{z\nu}$ as $g\rightarrow g_c^+$. From Eq.~\eqref{eq:MLRsq}, the linear response $\langle M_x\rangle$ upon a perturbation $\delta h$ will allow to resolve $\Delta_{s,N}(g)$ as well as $\Delta_N(g)$. Now the lowest frequency component of $S(\omega)$ is placed at $\omega_m=\Delta_{s,N}(g)$. Exact numerical simulation with $\delta h=10^{-3}$ and $g_0=1$ reveal the location of $\Delta_{s,N}(g_c)$ with good accuracy~\cite{SM}, which also allows to extract the dynamical critical exponent $z=1$ since $\Delta_{s,N}(g_c)\approx \pi/(2 N)\approx 8\Delta_{N}(g_c)$ for $N\gg 1$, as seen in Figs.~2(a) and 2(c) for various values of $N$. Finally, we consider a transverse-field Ising model with a long-range interactions, $H_{\rm long}=\sum_{\substack{i,j=1 (i<j)}}^{N}J_{i,j}\sigma^z_i\sigma^z_j+Jg\sum_{j=1}^N\sigma_j^x$, with $J_{i,j}=J|i-j|^{-r}$ for $i\neq j$ and $r>0$, which can realized in a trapped-ion platform~\cite{Kim:10,Islam:13,Richerme:14,Jurcevic:17}. As reported in~Ref.~\cite{Puebla:19}, for antiferromagnetic couplings $J<0$, the critical point for $r=2$ takes place at $g_c\approx 2.52$ with a corresponding $z=0.50(1)$. Assuming that  linear response at criticality is independent of the microscopic details of a system, we apply our method to determine the dynamical critical exponent in such non-integrable case. Upon numerically evaluating the linear response of $\langle M_x\rangle$ following the perturbation $g_c\rightarrow g_c+\delta g$ with $\delta g=0.01$, we can determine the energy spectrum [cf. Fig.~2(b)] and infer $z=0.47(2)$ for up to $N= 12$ spins [see Fig.~2(c)]~\cite{SM}. The good agreement with the expected value of $z$ demonstrates the power of the proposed approach.

\noindent
{\zz Conclusions.}
We have studied the quantum critical properties of  quantum many-body system using the framework of linear-response theory. We have shown that dynamical critical exponents  can be precisely determined from the linear response after a quench when combined with finite-size scaling arguments. We have illustrated our results with the transverse field Ising model and a long-range Kitaev chain, two integrable systems, as well  as with non-integrable models and non-zero temperature initial states. Our findings reveal an intimate correspondence between linear response theory and  quantum critical behavior of quantum many-particle systems. They moreover provide an accessible method to experimentally determine their dynamical critical exponents.

%
%

\section*{Acknowledgements}
\noindent A.B. was supported by H2020 through the MSCA IF pERFEcTO (Grant Nr.~795782) and the German Science Foundation DFG (Project Nr.~BR5221/4-1). R.P and M.P. acknowledge support from the DfE-SFI Investigator Programme (Grant ~5/IA/2864), the H2020-FETOPEN-2018-2020 project TEQ (Grant Nr.~766900),
the Royal Society Wolfson Research Fellowship
(RSWF\textbackslash R3\textbackslash183013),
the Royal Society International Exchanges Programme
(IEC\textbackslash R2\textbackslash192220),
the Leverhulme Trust Research Project Grant (Grant Nr.~RGP-2018-266), and the UK EPSRC.  This work was supported by a research grant from the Department for the Economy Northern Ireland under the US-Ireland R\&D Partnership Programme.
G.G acknowledge support from the Leverhulme Trust (RPG-2016-046) and the Spanish Agencia Estatal de Investigaci\'{o}n (Project~PID2019-107609GB-I00),  the QuantERA Grant C'MON-QSENS!, the Spanish MICINN PCI2019-111869-2.







\widetext
\clearpage
\begin{center}
\textbf{\large Supplemental Material:\\ \vspace{0.2cm} Linear-response approach to critical quantum many-body systems}
\end{center}
\setcounter{equation}{0}

\setcounter{figure}{0}
\setcounter{table}{0}
\setcounter{page}{1}
\makeatletter
\renewcommand{\theequation}{S\arabic{equation}}
\renewcommand{\thefigure}{S\arabic{figure}}
\renewcommand{\bibnumfmt}[1]{[S#1]}
\renewcommand{\citenumfont}[1]{S#1}

\begin{center}
  Ricardo Puebla$^{1,2}$, Alessio Belenchia$^{3,2}$, Giulio Gasbarri$^{4,5}$, Eric Lutz$^6$, and Mauro Paternostro$^{2}$ \\ \vspace{0.2cm} \small{
  {\em $^1$Instituto de F\'isica Fundamental, IFF-CSIC, Calle Serrano 113b, 28006 Madrid, Spain}\\
{\em $^2$Centre for Theoretical Atomic, Molecular, and Optical Physics,\\School of Mathematics and Physics, Queen’s University, Belfast BT7 1NN, United Kingdom}\\
{\em $^3$Institut f\"{u}r Theoretische Physik, Eberhard-Karls-Universit\"{a}t T\"{u}bingen, 72076 T\"{u}bingen, Germany}\\
 {\em $^4$Department of Physics and Astronomy, University of Southampton, Highfield Campus, SO17 1BJ, United Kingdom}\\
{\em $^5$F\'isica Te\`orica: Informaci\'o i Fen\`omens Qu\`antics, Department de F\'isica,\\Universitat Aut\`onoma de Barcelona, 08193 Bellaterra (Barcelona), Spain}\\
  {\em $^6$Institute for Theoretical Physics I, University of Stuttgart, D-70550 Stuttgart, Germany}}
\end{center}

\section{I. Linear response for a single spin}
We here evaluate the linear response of an observable $B$ for  the case of a single spin-1/2  with Hamiltonian $H_0 = \sigma_z$ and  perturbation $H_1=\sum_m c_m(t)\sigma_m$, where $\sigma_m$ are the  Pauli operators. Its density operator $\rho$ satisfies  
$\dot{\rho}=-i[\sigma_z,\rho]+\mathcal{L}_{1}(\rho)$,
with the  unitary perturbation $\mathcal{L}_{1}(\rho)=-i[H_1,\rho]$. The steady state of the unperturbed dynamics is  of the general form $\rho_0=(\1+f_z\sigma_z)/2$  with parameters $f_z$. A generic quantum observable can be further  written as a linear combination  ${B}=\sum_j b_j\sigma_j$ with coefficients $b_j$. 
According to  Eq.~\eqref{firstform}, the linear response of $B$ upon a time $t$  is explicitly given by
\begin{align}\label{unitarydephasing}
 \langle {B}\rangle&=f_z b_z+\int_0^t ds 2f_z\left[-(b_x c_y(s)+b_y c_x(s))\cos(2(t-s))+\sin(2(t-s))\left(c_x(s)b_x+c_y(s)b_y\right)\right].
\end{align}

\section{II. Ising model: comparison between linear response and exact dynamics}
\begin{figure}[t!]
\centering
\includegraphics[width=1\linewidth]{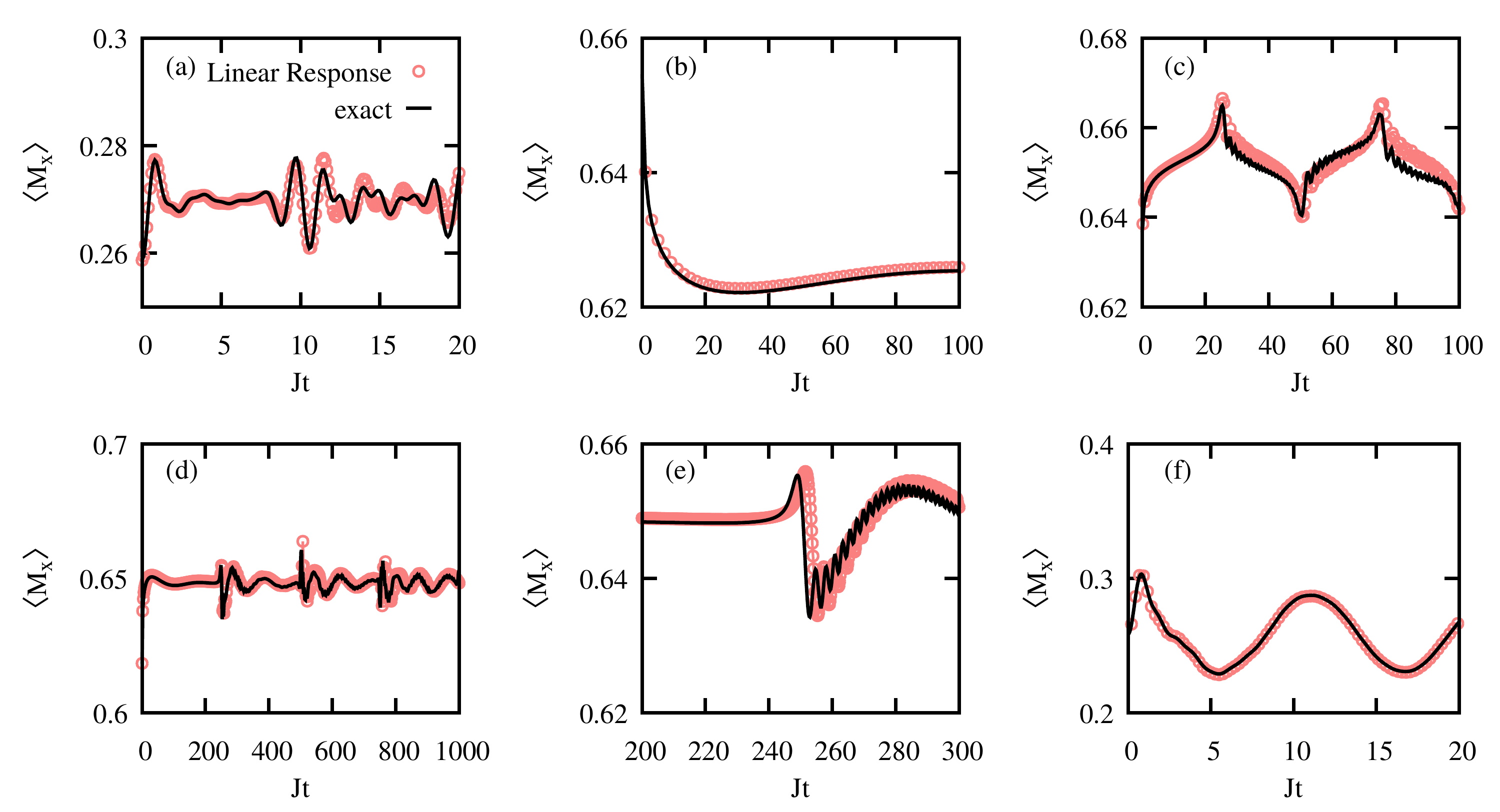}
\caption{\small{Comparison between the linear response (red circles) and the exact dynamics (solid black line) of $M_x$ for TFIM, initially in its ground state. Panel (a) shows the short-time behavior for $N=20$ spins and $g_0=0.5$ with $\delta g=0.02$, while panel (b) correspond to $N=1000$ spins crossing the QPT, i.e.  $g_0=1.01$  and $\delta g=-0.02$. The dynamics when starting at the critical point, $g_0=1$ and $\delta g=0.01$, is shown in panel (c) for $N=100$ spins. Panels (d) and (e) show the long-time dynamics for $N=1000$ spins with $g_0=0.99$ and $\delta g=0.02$. Note that panel (e) shows a zoom in the region $200\leq Jt\leq 300$.   In (f) the dynamics corresponds to a periodically-driven TFIM for $N=500$ spins and $g_0=0.5J$, $\delta g=0.05J\cos(2\omega_dt)$ with $\omega_d=0.28J$.}}
\label{figSM1}
\end{figure}

Let us start once again by considering the observable $M_x=\frac{1}{N}\sum_{j=1}^N\sigma_j^x$ for the TFIM. We have seen that, for a time-independent perturbation the linear response of this observable is given by
\begin{align}
  \langle M_x\rangle&=\frac{2}{N}\sum_{k>0}\left[\cos (2\theta^{(g_0)}_k)+\frac{4J\delta g\sin^2(2\theta^{(g_0)}_k)}{\epsilon_k(g_0)}\sin^2 (\epsilon_k(g_0)t) \right].
\end{align}
The previous expression can be computed in the thermodynamic limit $N\rightarrow \infty$, by taking the continuous limit,
\begin{align}
\lim_{N\rightarrow \infty}\langle M_x\rangle=\frac{1}{\pi}\int_0^\pi dk \left[\cos (2\theta^{(g_0)}_k)+\frac{4J\delta g\sin^2(2\theta^{(g_0)}_k)}{\epsilon_k(g_0)}\sin^2 (\epsilon_k(g_0)t) \right].
  \end{align}
Some examples are shown in Fig.~\ref{figSM1}, which demonstrate the very good agreement between the predictions of the linear response and the exact dynamics.

We can also consider the more general case of a time-dependent perturbation, e.g. $\delta g(t)=\delta g\cos(2\omega_d t)$. For a non-resonant frequency, $\omega_d\neq \epsilon_k(g_0)\ \forall k$, it follows
\begin{align}
  \langle M_{x}\rangle&=\frac{2}{N}\sum_{k>0}\left[\cos (2\theta^{(g_0)}_k)+4J\delta g\sin^2(2\theta^{(g_0)}_k)\frac{\epsilon_k(g_0)(\cos(2\omega_d t)-\cos(2\epsilon_{k}(g_0)t)}{2(\epsilon_k^2(g_0)-\omega_d^2)}\right].
\end{align}
In case the frequency matches $\epsilon_{k'}(g_0)$ for some $k'$, then ($\omega_d=\epsilon_{k'}(g_0)$), we obtain
\begin{align}
  \langle M_{x}\rangle&=\frac{2}{N}\sum_{\substack{k>0\\k\neq k'}}\left[\cos (2\theta^{(g_0)}_k)+4J\delta g\sin^2(2\theta^{(g_0)}_k)\frac{\epsilon_k(g_0)(\cos(2\omega_d t)-\cos(2\epsilon_{k}(g_0)t)}{2(\epsilon_k^2(g_0)-\omega_d^2)}\right]\nonumber \\
  &+\frac{2}{N}\left[\cos (2\theta^{(g_0)}_{k'})+4J\delta g\sin^2(2\theta^{(g_0)}_{k'})\frac{t\sin (2\epsilon_{k'}(g_0)t)}{2}\right].
\end{align}
An example illustrating the good agreement between the prediction of the linear response for a periodically-perturbed Ising model and its exact dynamics is shown in Fig.~\ref{figSM1}(f).

Finally, we can focus on a different observable, as for example the two-point correlation function of the order parameter  $M_{zz}=\frac{1}{N}\sum_{j=1}^N\sigma_j^z\sigma_{j+1}^z$, with $\sigma_{N+1}^z=\sigma_1^z$. Again, we assume the initial state to be the ground state at $g_0$, i.e. $f^z_k=-1\ \forall k$. The observable $M_{zz}$ in the $k$-momentum subspace reads as 
\begin{align}
M_{zz}=\frac{1}{N}\sum_{n=1}^N\sigma_n^z\sigma_{n+1}^z\rightarrow M_{zz}=\sum_{k>0}\Psi_k^\dagger M_{zz,k} \Psi_k, \ {\rm with} \  M_{zz,k}=\frac{2}{N}[\cos k b \sigma_z^k-\sin k b \sigma_x^k]
  \end{align}
In the rotated basis (i.e. eigenbasis in the $k$-subspace), we find
\begin{align}
    M_{zz}&=\frac{2}{N}[ \cos k b (\cos(2\theta^{(g_0)}_k)\tilde{\sigma}_k^z-\sin(2\theta^{(g_0)}_k)\tilde{\sigma}_k^x)+\sin k b(\sin(2\theta^{(g_0)}_k)\tilde{\sigma}_k^z+\cos (2\theta^{(g_0)}_k)\tilde{\sigma}_k^x)]\\
    &=\frac{2}{N}[\tilde{\sigma}_k^z(\cos k b \cos(2\theta^{(g_0)}_k)+\sin k b\sin(2\theta^{(g_0)}_k)+\tilde{\sigma}_k^x(-\cos k b \sin(2\theta^{(g_0)}_k)+\sin k b \cos(2\theta^{(g_0)}_k))].
\end{align}
Then, it is straightforward to find the corresponding expression for the linear response of this observable, which reads as
\begin{align}
    \langle M_{zz}\rangle=\frac{2}{N}\sum_{k>0}&\left[-(\cos k b\cos(2\theta^{(g_0)}_k)+\sin k b\sin(2 \theta^{(g_0)}_k))\right.\\&\left.+\frac{4J\sin(2 \theta^{(g_0)}_k)(\sin k b\cos (2\theta^{(g_0)}_k)-\cos k b \sin(2\theta^{(g_0)}_k))}{\epsilon_k(g_0)}\int_0^{\epsilon_k(g_0)t} ds \delta g(s) \sin(2(\epsilon_{k}(g_0)t-s)) \right]
\end{align}
For a time-independent perturbation, the previous expression simplifies to
\begin{align}
    \langle M_{zz}\rangle=\frac{2}{N}\sum_{k>0}&\left[-(\cos k b\cos(2\theta^{(g_0)}_k)+\sin k b\sin(2 \theta^{(g_0)}_k))\right.\\&\left.+4J(\delta g)\sin(2 \theta^{(g_0)}_k)(\sin k b\cos(2\theta^{(g_0)}_k)-\cos k b \sin(2\theta^{(g_0)}_k))\frac{\sin^2(\epsilon_k(g_0)t)}{\epsilon_k(g_0)} \right].
\end{align}
The results of this analysis are very similar to those shown in Fig.~\ref{figSM1}, and again, such observable would allow for the determination of the critical exponents of the many-body system.

\section{III. LRK chain model}
Let us now consider the case of the LRK chain of fermionic particles~\cite{Kitaev:01SM,Vodola:14SM,Alecce:17SM}. The Hamiltonian reads
\begin{align}
    H_{\rm LRK}=-\sum_j \left[\sum_{r>0} J\left(J_r(c_j^\dagger c_{j+r}+c_{j+r}^\dagger c_j)+ d_r ( c_j c_{j+r}+c_{j+r}^\dagger c_j^\dagger )\right)- \mu\left( c_j^\dagger c_j-\frac{1}{2}\right) \right].
\end{align}
where, $\mu$ is the chemical potential which controls the QPTs appearing in this model, and
\begin{align}
J_r^\alpha=\frac{1}{N_\alpha \overline{r}^\alpha}, \quad d_r^\beta=\frac{1}{N_\beta \overline{r}^\beta}
  \end{align}
with $\overline{r}=\min(r,N/2-r)$ as we take periodic boundary conditions.

\begin{figure}[t!]
\centering
\includegraphics[width=1\linewidth]{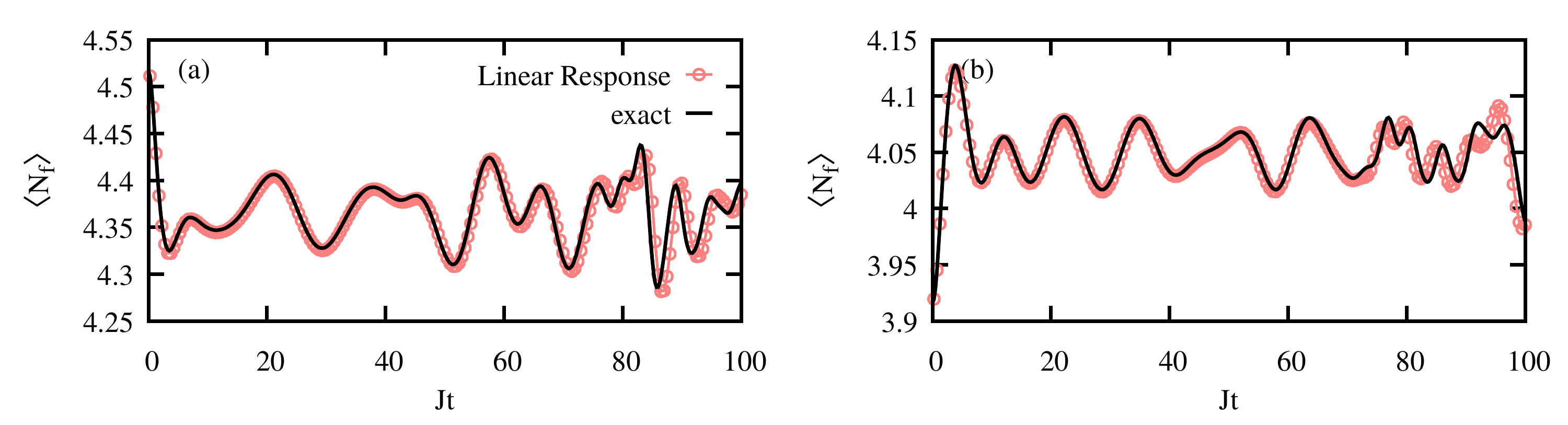}
\caption{\small{Comparison between the linear response (red circles) and the exact dynamics (solid black line) of the number of fermions $N_f$ for a perturbed long-range Kitaev chain initialized in the ground state. In both cases, $\mu_0=2J$, $\alpha=5/2$, $N=100$, while $\delta \mu=0.01J$ and $\beta=3/2$ in (a) and $\delta\mu=-0.01J$ and $\beta=5/4$ in (b).}}
\label{figSM2}
\end{figure}

By Fourier-transforming the fermionic operators, we find a Block diagonal structure of the Hamiltonian, 
\begin{align}
    H_{\rm LRK}= \sum_{k>0}\Psi_k^\dagger H_k \Psi_k
\end{align}
with $\Psi_k^\dagger=(c_k^\dagger,c_{-k})$ and $H_{k}=h_z(\mu)\sigma_z^k+h_x\sigma_x^k$ where now $h_z(\mu)=\mu/2-2J\sum_{r>0}J_r\cos(krb)/N_\alpha$ and $h_x=-J \sum_{r>0}d_r\sin (krb)/N_\beta$, so that upon a diagonalization one finally obtains $H_{0,k}=\epsilon_k(\mu)\tilde{\sigma}_z^k$ with $\epsilon_k(\mu)=\sqrt{h_z^2(\mu)+h_x^2}$, and $\theta^{(\mu)}_k$ the mixing angle as given for the TFIM. The perturbation in the chemical potential $\delta \mu(t)$ leads to $H_{1}=\delta\mu(t)/2 \sigma_z^k$. Hence, the underlying structure is very similar to the TFIM. The observable $N_f=\sum_{n=1}^N c_n^\dagger c_n$ can be expressed as $N_f=\sum_{k>0}\Psi_k^\dagger \sigma_z^k\Psi_k$ in the Fourier-transformed fermionic operators. As explained above for the TFIM, a direct substitution in Eq.~\eqref{unitarydephasing} leads to 
\begin{align}
    \langle N_f \rangle=\sum_{k>0}\left[-\cos 2\theta^{(\mu_0)}_k-\frac{2(\delta \mu) \sin^2(2\theta^{(\mu_0)}_k)}{\epsilon_k(\mu_0)}\sin^2(\epsilon_k(\mu_0)t)\right],
\end{align}
In Fig.~\ref{figSM2} we show the comparison between the exact dynamics and the linear response, which show an excellent agreement. 

\section{IV. Non-zero temperature initial states}
As commented in the main text, the method based on the linear response of the a many-body system close to the critical point works also for non-zero temperature initial states. In this case, each of the fermions becomes excited with a certain probability so that the initial thermal equilibrium state of $H_{0,k}$ is $\rho_0^k=(\1+f_z^k\tilde{\sigma}_z^k)/2$ with $f_z^k=-\tanh(\epsilon_k(g_0)/(k_b T))$, and both parity subspaces must be taken into account. We note, however, that since $H_{\rm TFIM}$ conserves the parity symmetry $Z_2$, the energy gap $\Delta_{s,N}(g)$ cannot be resolved with non-zero temperature initial states (see {\em non-integrable models} in the main text). Hence, the lowest frequencies for $S(\omega)$ take place at $\omega_m=2\epsilon_{k_1}(g_0)$, $2\epsilon_{k=2\pi/N}(g_0)$ and $2\epsilon_{k_2}(g_0)$. Note that $k=2\pi/N$ corresponds to the other parity subspace. For increasing temperature the Fourier components are reduced, as commented in the main text. 

\section{V. Longitudinal and long-range transverse-field Ising models}
A perturbation to the TFIM with a longitudinal magnetic field, i.e. according to
\begin{align}
  H_\text{longitudinal}=H_{\rm TFIM}+Jh \sum_{j=1}^{N}\sigma_j^z,
\end{align}
breaks its parity symmetry and integrability. Upon a perturbation $h_0=0\rightarrow \delta h$, an initially prepared ground state of $H_{\rm TFIM}$ at $g_0$ will tunnel to the other parity subspace. The linear response allows to determine the energy spectrum at $g_0$, but in this case also the excitation energies among subspaces with distinct parity. In particular, the energy gap for the ground state with opposite parity is given by $\Delta_{s,N}(g)=E_0^--E_0$ with $E_0^-=-2J(1+\sum_{n=1}^{N/2-1}\sqrt{g^2+1-2g\cos(2n\pi/N)}$. For $g<g_c$, this energy separation vanishes exponentially, while at the critical point enables the determination of the dynamical critical exponent $z$ (cf. Fig.~2(a) and (c) of the main text). Indeed, a fit to the obtained lowest-frequency components leads to $z=1.01(1)$, in agreement with the theoretical value $z=1$.

Finally, we show the results for a transverse-field Ising model with long-range interactions, given by the Hamiltonian
\begin{align}
  H_{\rm long}=\sum_{\substack{i,j=1\\i<j}}^N J_{i,j}\sigma_i^z\sigma_j^z+J g\sum_{j=1}^N\sigma_j^x,
  \end{align}
with $J_{i,j}=J|i-j|^{-r}$ with $i\neq j$. This model features quantum phase transitions, whose critical exponents depend on the range of the interactions, i.e. on the exponent $r$. For $r=2$, $J<0$, it has been reported in~\cite{Puebla:19SM} that the critical point takes place at $g_c=2.52$, with a dynamical critical exponent $z=0.50(1)$. Proceeding as before, we find $z=0.47(2)$ for sizes up to $N=12$ spins (cf. Fig.~2(b) and (c) of the main text).

\end{document}